\documentclass[12pt]{article}
\usepackage{pdproc,graphicx} 

\begin{document}

\renewcommand{\thefootnote}{\alph{footnote}}
  
\title{
 WHAT WE (WOULD LIKE TO) KNOW\\[1mm] ABOUT THE NEUTRINO MASS}

\author{G.L.~FOGLI~$^{1,2*}$, E.~LISI~$^{2*}$, A.~MARRONE~$^{1,2}$, 
A.~PALAZZO~$^3$,
A.M.~ROTUNNO~$^{1,2}$}
\smallskip
\address{ $^1$~Dipartimento di Fisica, Universit\`a di Bari,
Via Amendola n.176, 70126 Bari, Italy \\[+0.9mm]
$^2$~Sezione INFN di Bari, Via Orabona n.4, 70126 Bari, Italy\\[+0.4mm]
$^3$~AHEP Group, Institut de F\'isica Corpuscular, CSIC/Universitat de Val\`encia,\\
Edifici Instituts d'Investigaci\'o, Apt.\ 22085, 46071 Val\`encia, Spain }

\centerline{\footnotesize $^*$Speaker. \tt fogli@ba.infn.it}
\abstract{We present updated values for the mass-mixing parameters relevant
to neutrino oscillations, with particular attention to emerging hints
in favor of $\theta_{13}>0$. We also discuss the status of absolute 
neutrino mass observables, and a possible approach to constrain
theoretical uncertainties in neutrinoless double beta decay.
Desiderata for all these issues are also briefly mentioned.}

\normalsize\baselineskip=15pt


\section{Neutrino oscillation parameters and hints of $\theta_{13}>0$}

In the last decade, a series of beautiful $\nu$ oscillation experiments, interpreted within
a theoretical framework with three massive and mixed neutrinos, has provided
stringent constraints on the $\nu$ mass-squared differences
($\delta m^2,\,\Delta m^2$) and mixing angles $(\sin^2\theta_{12},\,\sin^2\theta_{23},\,\sin^2\theta_{13})$---see  \cite{Foglireview} for notation and conventions. Desiderata 
include the sign of $\Delta m^2$ (i.e., the $\nu$
mass spectrum hierarchy), a possible CP-violating phase $\delta$, and
a lower bound (if any) on the smallest $\nu$ mixing angle $\theta_{13}$.

At this NO-VE Workshop, we have presented recent progress on the latter issue,
by showing how the combination of recent solar and long-baseline
reactor (KamLAND) data leads to 
a slight preference for $\sin^2\theta_{13}\sim 10^{-2}$.
More precisely, using solar $\nu$ data
{\em prior\/} to the Neutrino~2008 Conference \cite{Nu2008}, we have
noted that the slight difference between the value of $\sin^2\theta_{12}$
preferred by such data, as compared with the one preferred by KamLAND data,
can be significantly reduced for $\theta_{13}>0$---see the panels in Fig.~1. 
This reduction, originating from different functional forms
of the survival probability $P_{\nu_e\to\nu_e}(\sin^2\theta_{12},\sin^2\theta_{13})$ in the two
classes of experiments, led to a slight preference for $\theta_{13}>0$
(at the level of $\sim\!\!0.5\sigma$).
An older, independent preference for $\theta_{13}>0$ 
from atmospheric neutrino data had already been found in \cite{Foglireview}
(at the level of $\sim\!\!0.9\sigma$).
Added together, 
these hints provided a $1\sigma$ indication in favor of $\theta_{13}>0$: 
$\sin^2\theta_{13}\simeq 0.01\pm 0.01$, as presented at this Workshop
using available data. We then remarked that new solar $\nu$
data could potentially corroborate such indication.

After Neutrino~2008 \cite{Nu2008}, we have
performed a follow-up global analysis with the latest (SNO-III) solar neutrino results
\cite{theta13}. 
The analysis has indeed sharpened such intriguing
indication, which now reaches an interesting C.L.\ of  
$\sim\!90\%$ (i.e., $\sim\!\!1.6\sigma$):
\begin{equation}
\sin^2\theta_{13}\simeq 0.016\pm 0.010\ .
\end{equation}

\begin{figure}[t]
\begin{minipage}{17.5pc}
\includegraphics[height=2.2in,width=7cm]{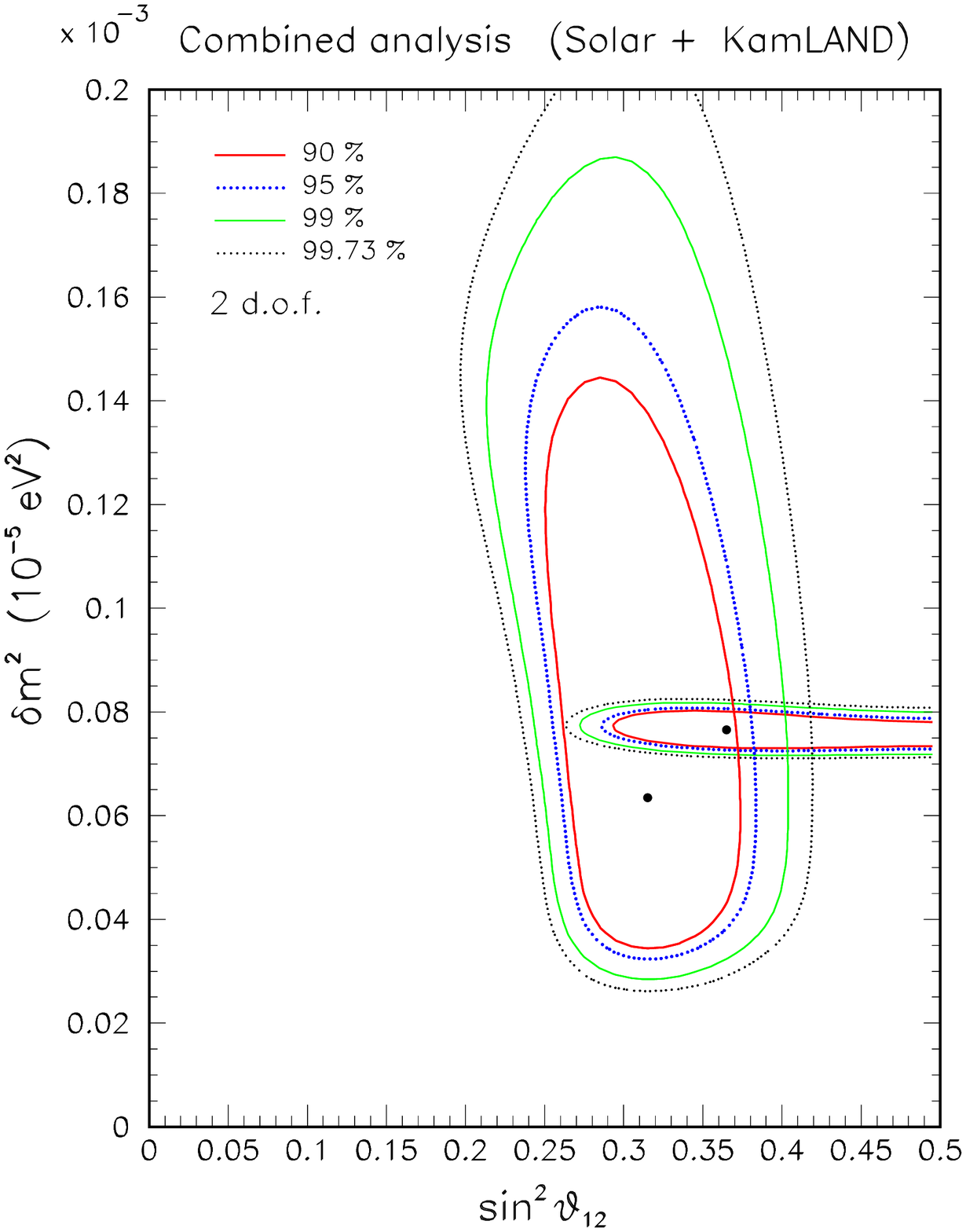}
\end{minipage}
\hspace{1pc}%
\begin{minipage}{17.5pc}
\includegraphics[height=2.2in,width=7cm]{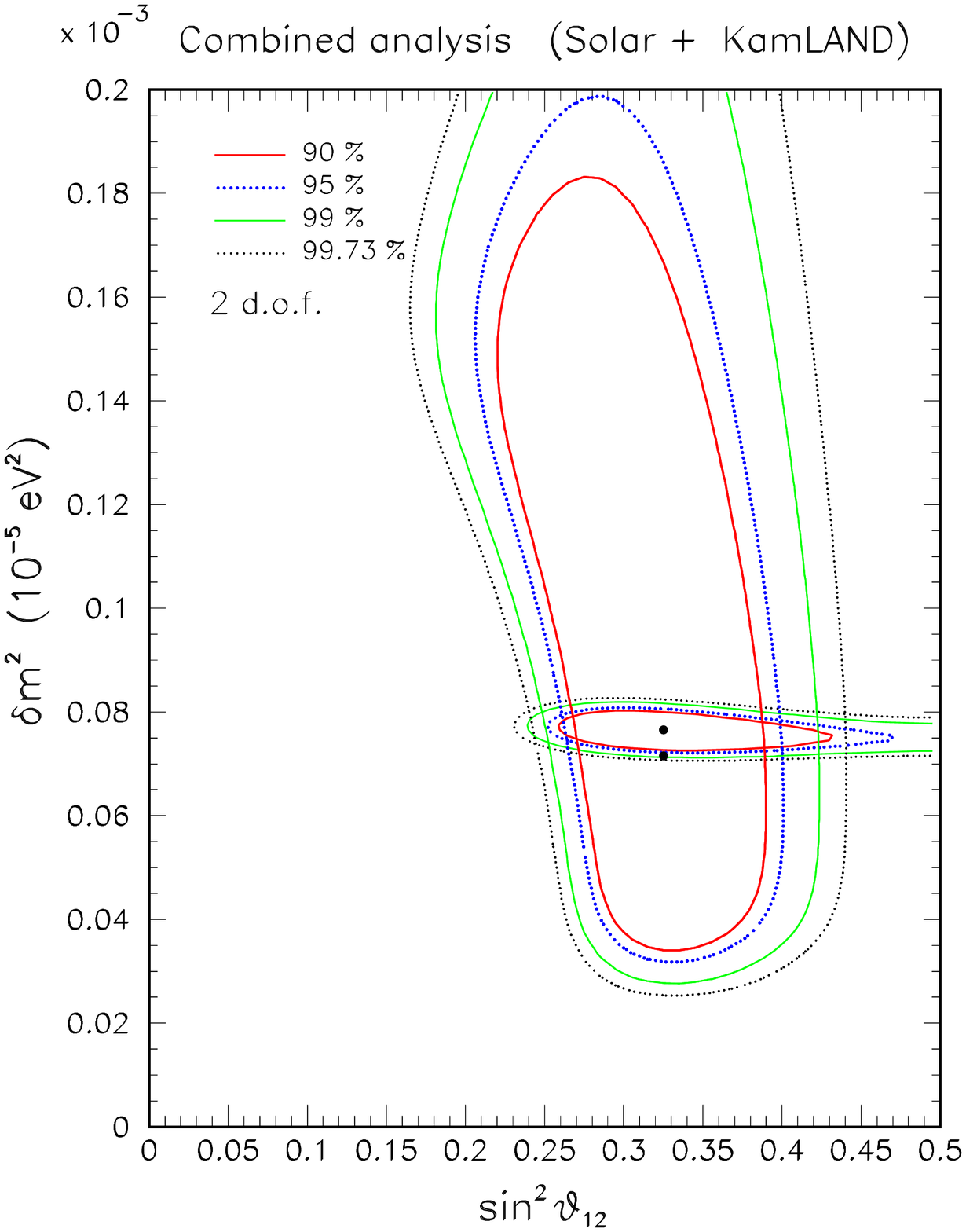}
\end{minipage} 
\caption{Comparison of regions allowed by KamLAND data (2008)
and by solar $\nu$ data (prior to Neutrino 2008), according to 
our analysis. Contours refer to two dof. Black dots
represent best-fit points. The left panel refers to $\sin^2\theta_{13}=0$
(fixed); note the slight distance between the best-fit points. The right
panel refers to $\sin^2\theta_{13}=0.03$ (fixed); the best-fit points
are closer, although the overall goodness of the fit (not reported here)
is slightly worsened.\vspace*{-2mm}}
\end{figure}

Figure~2 (left panel) shows the regions separately allowed
at $1\sigma$ ($\Delta\chi^2=1$, dotted) and $2\sigma$ ($\Delta\chi^2=4$, solid)
from the analysis of solar (S) and KamLAND (K)  neutrino data, in the
plane spanned by $(\sin^2\theta_{12},\, \sin^2\theta_{13})$.
These parameters are 
positively and negatively correlated in the S and K regions, respectively,
as a result of different functional forms for $P_{ee}(\sin^2\theta_{12},\sin^2\theta_{13})$
in the two cases.
The S and K allowed regions, which do not overlap at $1\sigma$
for $\sin^2 \theta_{13}=0$, merge for $\sin^2\theta_{13}\sim\mathrm{few}\times 10^{-2}$. 
The best fit (dot) and error ellipses (in black) for the S+K  combination are shown
in the middle panel of Fig.~2; a hint of $\theta_{13}>0$ emerges at  $\sim\!\!1.2\sigma$ level. Finally,
the  independent ($\sim \! 0.9\sigma$)
hint of $\theta_{13}>0$ from the 
combination of atmospheric, LBL accelerator, 
and CHOOZ data reinforces the overall preference for $\theta_{13}>0$, which emerges at the overall
level of $\sim\!\! 1.6\sigma$ in the right panel of Fig.~2 (all data).

\begin{figure}[b]
\center
\vspace*{-2mm}
\hspace*{-6mm}
\includegraphics[height=2.0in,width=14.5cm]{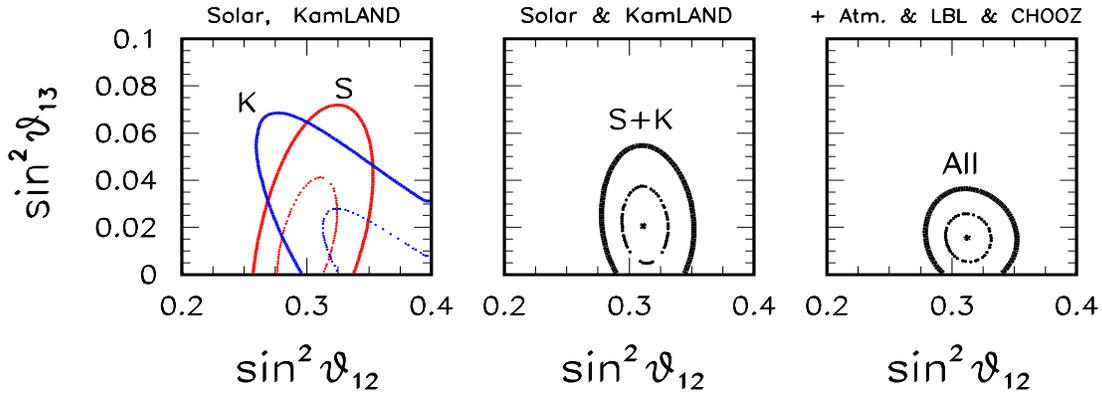}
\vspace*{-1mm}
\caption{Analysis after Neutrino~2008: Contours of the regions allowed at $1\sigma$ (dotted)
and $2\sigma$ (solid) in the plane $(\sin^2\theta_{12},\,\sin^2\theta_{13})$. Left and middle
panels: solar (S) and KamLAND (K) data, both separately (left)
and in combination (middle). Right panel: All data.
Taken from Ref.~\protect\cite{theta13}.}
\end{figure}
\newpage

\begin{figure}[t]
\center
\hspace*{-2mm}
\includegraphics[height=2.0in,width=15.cm]{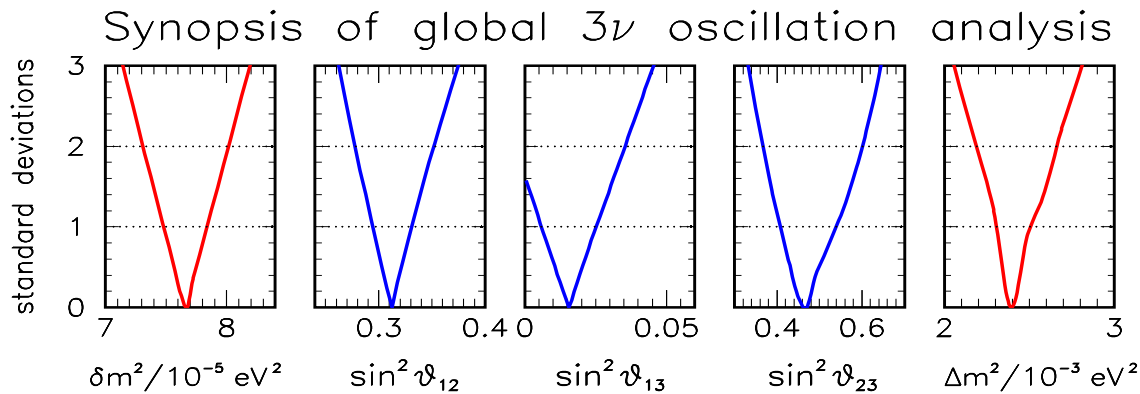}
\vspace*{-1mm}
\caption{Global analysis after Neutrino~2008,
from Ref.~\protect\cite{Addendum}:  Bounds on the 
mass-mixing oscillation parameters, in terms of standard deviations  from 
the best fit. Note again the $1.6\sigma$ preference for $\theta_{13}>0$.
Desiderata include a possible confirmation of such preference in future
reactor and accelerator oscillation searches.}
\end{figure}
\newpage

At this Workshop we also updated the estimates of all 
the mass-mixing parameters. Here we present the latest results
including Neutrino~2008 data \cite{Addendum}.
Figure~3 displays a synopsis of the $\nu$
parameters, in terms of
standard deviations $n_\sigma$ from the best fit ($n_\sigma=\sqrt{\Delta\chi^2}$ after $\chi^2$ marginalization).
Table~\ref{Synopsis} summarizes the numerical ranges.
Note that in the the discussion of Sec.~2
we shall show results at a conservative $2\sigma$ (95\%) C.L.,
 in which case only an upper bound can be placed on $\theta_{13}$.

\begin{table}[h]
\caption{\label{Synopsis} Global $3\nu$ oscillation analysis (2008): best-fit values and
allowed $n_\sigma$ ranges, from Ref.~\protect\cite{Addendum}.}
\resizebox{\textwidth}{!}{
\begin{tabular}{cccccc}
\hline
\hline
Parameter & $\delta m^2/10^{-5}\mathrm{\ eV}^2$ & $\sin^2\theta_{12}$ & $\sin^2\theta_{13}$ & $\sin^2\theta_{23}$ &
$\Delta m^2/10^{-3}\mathrm{\ eV}^2$ \\[4pt]
\hline
Best fit        &     7.67     &  0.312          &  0.016          &  0.466          &  2.39 \\
$1\sigma$ range & 7.48~--~7.83 & 0.294~--~0.331  & 0.006~--~0.026  & 0.408~--~0.539  & 2.31~--~2.50 \\
$2\sigma$ range & 7.31~--~8.01 & 0.278~--~0.352  & $<0.036$        & 0.366~--~0.602  & 2.19~--~2.66 \\
$3\sigma$ range & 7.14~--~8.19 & 0.263~--~0.375  & $<0.046$        & 0.331~--~0.644  & 2.06~--~2.81 \\
\hline
\hline
\end{tabular}
}
\end{table}

\section{Status of absolute neutrino mass observables}

The three main
observables sensitive to absolute $\nu$ masses are: the effective mass
$m_\beta$ in single beta  decay, the effective Majorana mass $m_{\beta\beta}$ 
in neutrinoless double beta ($0\nu2\beta$) decay, and the sum of $\nu$ masses $\Sigma$ in cosmology---see Ref.~\cite{Foglireview} for notation. Desiderata include an undisputed nonzero signal for at
least one such quantity. 

\subsection{$0\nu2\beta$ decay updates}
The final analysis of part of the Heidelberg-Moscow (HM) Collaboration reports a $0\nu2\beta$ signal in $^{76}$Ge
with half-life $T^{0\nu}_{1/2}=2.23^{+0.44}_{-0.31}\times 10^{25}$~y ($1\sigma$ errors) at a claimed 
C.L.\ $>6\sigma$. Desiderata include an independent check of this claim
in a different experiment. Using generous uncertainties for the
$0\nu2\beta$ nuclear
matrix elements (NME) we find
\begin{equation}
\log (m_{\beta\beta}/\mathrm{eV}) = -0.54 \pm 0.26 \ \ \ (\mathrm{HM\ claim},\; 2\sigma)\ .
\label{HMclaim}
\end{equation}

The Cuoricino experiment does not find $0\nu2\beta$ decay signals  in $^{130}$Te and
quotes $T^{0\nu}_{1/2}>2.5\times 10^{24}$~y at 95\% C.L. Using  the
latter limit as $\log(T^{0\nu}_{1/2}/\mathrm{y})>24.4$, we get 
\begin{equation}
\log (m_{\beta\beta}/\mathrm{eV}) < [-0.63,\,-0.07] \ \ \ (\mathrm{Cuoricino},\; 2\sigma)\ ,
\end{equation}
where the range due to the $2\sigma$ uncertainty of the NME is explicitly reported.

A comparison of the corresponding $m_{\beta\beta}$ ranges ($2\sigma$),
\begin{eqnarray}
0.16 < m_{\beta\beta}/\mathrm{eV} < 0.52 &&  \ \ \ (\mathrm{HM\ claim})\ ,\label{HMclaim2}\\
0\leq  m_{\beta\beta}/\mathrm{eV} < 0.23 && \ \ \ (\mathrm{Cuoricino,\ ``favorable"\ NME})\ ,\\
0\leq  m_{\beta\beta}/\mathrm{eV} < 0.85 && \ \ \ (\mathrm{Cuoricino,\ ``unfavorable"\ NME})\ ,
\end{eqnarray}
shows that current Cuoricino data may or may not disfavor a fraction of the HM 
range for $m_{\beta\beta}$ at $2\sigma$, depending on the (still quite uncertain)
value of the $^{130}$Te $0\nu2\beta$ NME. Desiderata include a reduction of
the NME uncertainties (see also Sec.~3).
 Therefore, the $0\nu2\beta$ claim 
remains an open issue at present, and we shall consider the possibility that
it corresponds to a real signal.  See Ref.~\cite{Addendum} and references therein for details about the above limits.

\subsection{Cosmology updates} In Ref.~\cite{Addendum}, in collaboration 
with A.~Melchiorri, P.~Serra, J.~Silk,
 and A.~Slosar, we have also updated the constraints on $\Sigma$ including
WMAP 5-year and other data. We consider four representative combinations of cosmological data, which lead to 
increasingly stronger upper limits on $\Sigma$: (1)
CMB anisotropy data from: WMAP~5y,
ACBAR, VSA, CBI, and BOOMERANG; 
(2) the above CMB results plus the HST prior on the value of the reduced
  Hubble constant, and the luminosity
  distance SN-Ia data;
(3) The data in (2) plus BAO data;
(4) all the previous data, plus Ly$\alpha$ forest clouds data. 
The corresponding upper limits on $\Sigma$ are
summarized in Table~\ref{tableCASES}.
We shall focus on the two extreme cases 1 and 4.
\begin{table}[h]
\caption{\label{tableCASES} Cosmological datasets 
and corresponding $2\sigma$ bounds on $\Sigma=m_1+m_2+m_3$.}
\vspace*{2mm}
\begin{center}
\begin{tabular}{llr}
\hline
\hline
Case 	& Cosmological data set		& $\Sigma$ (at $2\sigma$)\\[4pt]
\hline
1 		& CMB  													& $<1.19$ eV \\
2		& CMB + HST + SN-Ia										& $<0.75$ eV \\ 
3 		& CMB + HST + SN-Ia + BAO 								& $<0.60$ eV \\ 
4		& CMB + HST + SN-Ia + BAO + Ly$\alpha$ 					& $<0.19$ eV\\
\hline
\hline
\end{tabular}
\end{center}
\end{table}

\subsection{Neutrinoless double beta decay versus cosmology plus oscillations}

Figure~4 (left) shows the regions allowed at $2\sigma$ in normal and inverted hierarchy (slanted bands)
by the combination of oscillation results with the first dataset in Table~\ref{tableCASES} (CMB), 
in the plane spanned by ($\Sigma,\,m_{\beta\beta}$). This is the most conservative
case, with the weakest limits on $\Sigma$, and the largest
overlap between the regions separately allowed by oscillation+CMB data and
by the $0\nu2\beta$ claim. The results of a global $\chi^2$ fit are shown as a thick black
wedge in the upper right part of the figure. Such global combination would correspond  to nearly degenerate masses in the range
$$
m_1\simeq m_2\simeq m_3 \in [0.15,\,0.46]~\mathrm{eV}\ \ (2\sigma)\ . 
$$

In this case (degenerate spectrum), the preferred range for effective neutrino mass in $\beta$ decay would also
be $m_\beta\in [0.15,\,0.46]$~eV.
In the upper half of this range, the KATRIN $\beta^-$ experiment could make a
$5\sigma$ discovery, according to the estimated sensitivity. A $3\sigma$ evidence
could still be found in KATRIN for $m_\beta\sim 0.3$~eV. Below this value, the sensitivity would
be rapidly degraded, and only upper bounds could be placed for $m_\beta< \sim0.2$~eV.

\begin{figure}[t]
\begin{minipage}{17.5pc}
\includegraphics[height=2.5in,width=7cm]{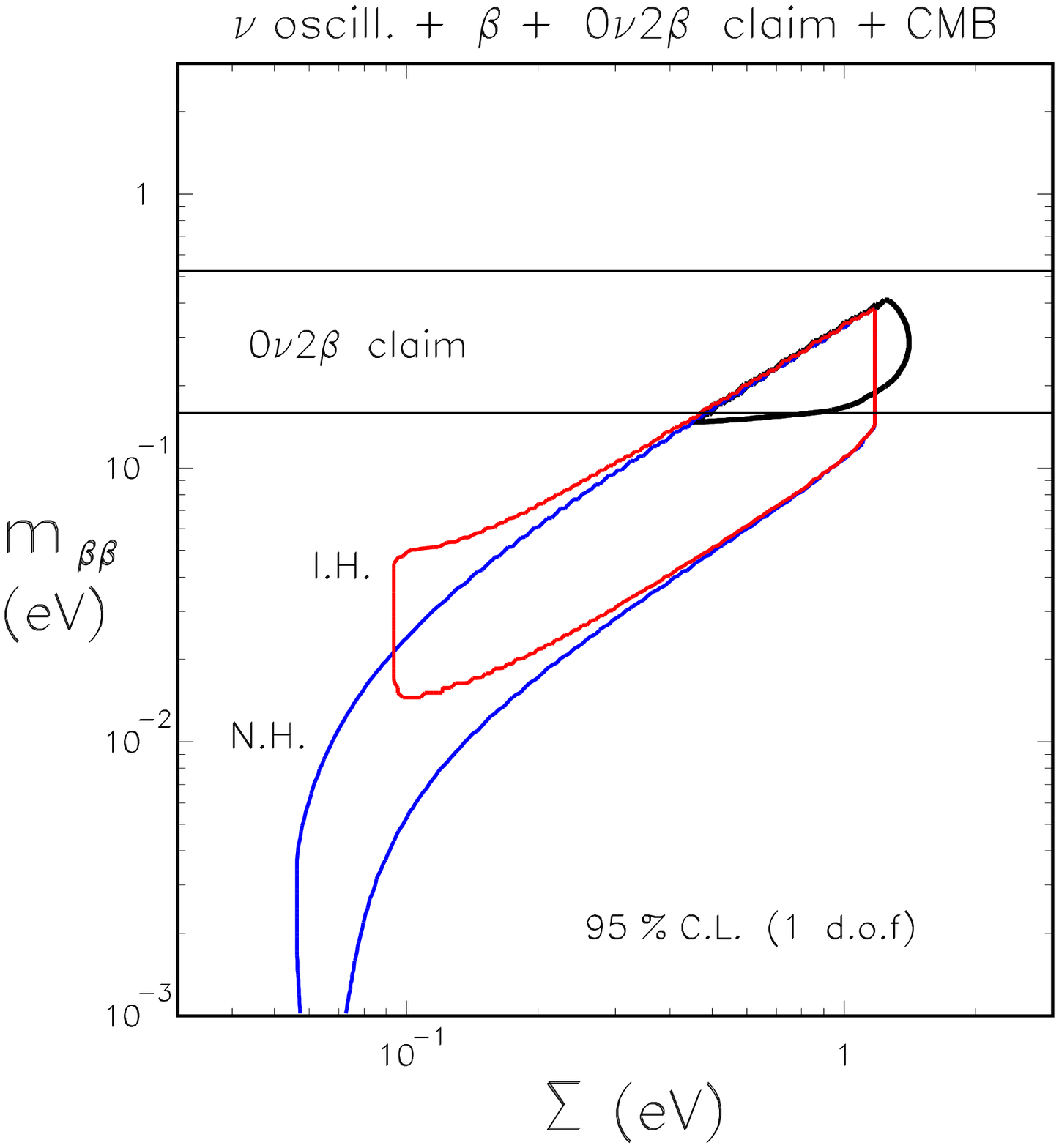}
\end{minipage}
\hspace{1pc}%
\begin{minipage}{17.5pc}
\includegraphics[height=2.5in,width=7cm]{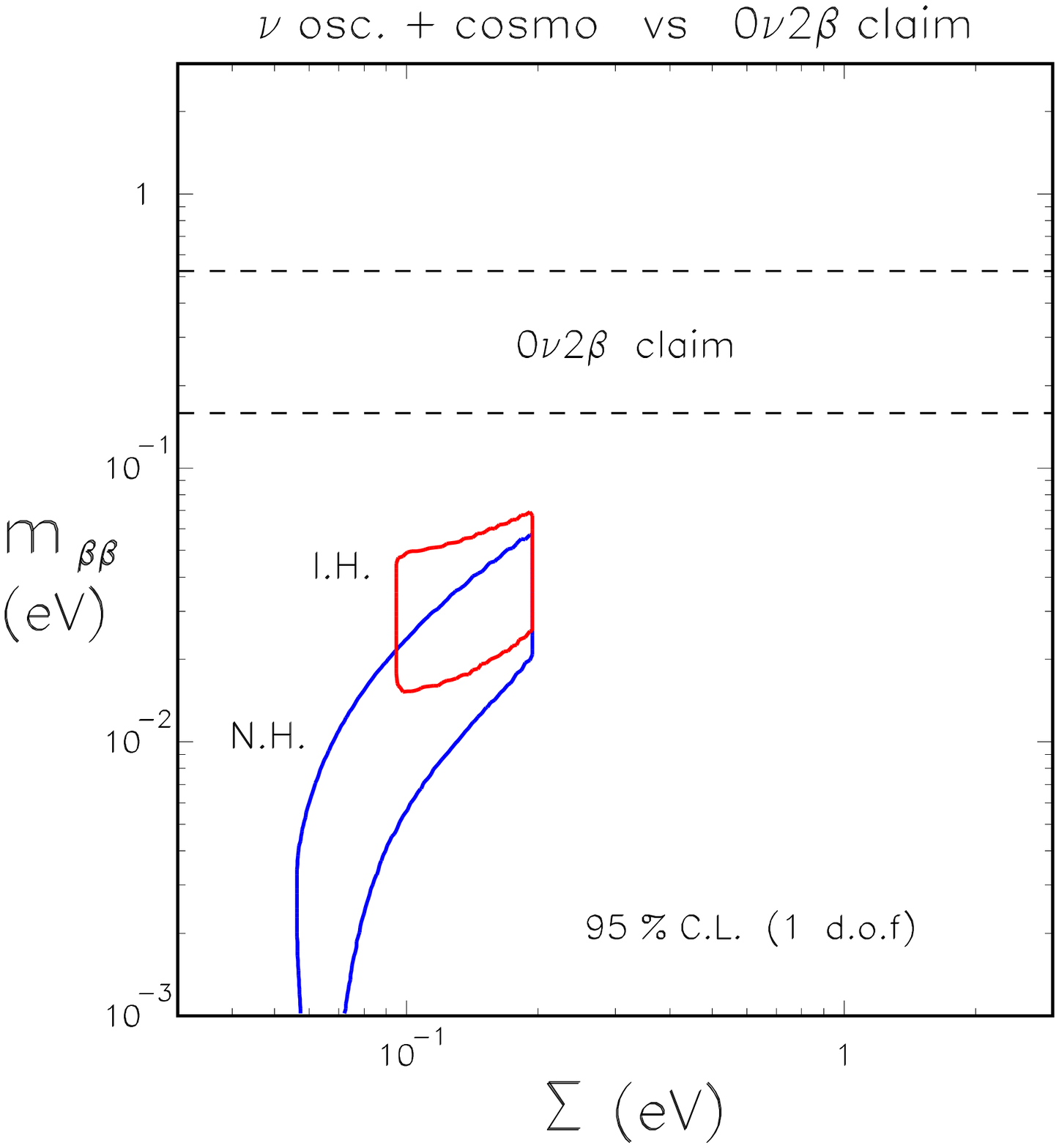}
\end{minipage} 
\caption{
Comparison on bounds placed by oscillation data (slanted bands,
for normal and inverted hierarchy), cosmological
data (vertical limits) and the $0\nu2\beta$ claim (horizontal band),
 in the plane $(\Sigma,\,m_{\beta\beta})$. Left panel:
Global combination of oscillation plus CMB data (case 1 in Table~\protect\ref{tableCASES}) with the
$0\nu2\beta$ decay claim. Right panel: Bounds from oscillation plus all cosmological data (case 4 in Table~\protect\ref{tableCASES}), contrasted with the $0\nu2\beta$ decay claim. Figure taken
from Ref.~\protect\cite{Addendum}.}
\vspace*{-6mm}
\end{figure}

The right panel in Fig.~3 is analogous to the left one, but refers to the 4th dataset in Table~\ref{tableCASES} (all cosmo
data, including Ly$\alpha$). In this case, the allowed regions do not overlap and cannot be combined, 
since the relatively strong cosmological limit $\Sigma<0.19$~eV
implies $m_{\beta\beta}< \sim 0.08$~eV, in contradiction with Eq.~(\ref{HMclaim2}).
Solutions to this discrepancy would require that either some data or their interpretation are wrong.
\clearpage
\newpage

\section{NME uncertainties in $0\nu2\beta$ decay}

Constraining NME uncertainties is crucial to compare results
from different $0\nu2\beta$ experiments and to provide well-defined
values (or limits) for $m_{\beta\beta}$, as shown in Sec.~2.1.
Here we report on an approach to this problem developed in Ref.~\cite{nme}
(in collaboration with A.~Faessler, V.~Rodin, and F.~Simkovic) which,
although currently limited to two nuclei, appears to be promising.

Estimates of nuclear matrix elements for $0\nu2\beta$ decay 
based on the quasiparticle random phase approximations (QRPA) are affected by 
theoretical uncertainties, which can be substantially reduced by fixing the 
unknown strength parameter $g_{pp}$ of the residual particle-particle interaction 
through one experimental constraint --- most notably through the two-neutrino double 
beta decay ($2\nu2\beta$) lifetime. However, it has been noted that the $g_{pp}$ 
adjustment via $2\nu2\beta$ data may bring QRPA models in disagreement
with independent data on electron capture (EC) and single beta decay ($\beta^-$) 
lifetimes. Actually, in two nuclei of interest for $0\nu2\beta$ decay
($^{100}$Mo and $^{116}$Cd), for which all such data are available, 
we have shown in Ref.~\cite{nme}
that the disagreement vanishes, provided that the axial vector coupling 
$g_A$ is treated as a free parameter, with allowance for $g_A<1$ (``strong quenching''). 
Three independent lifetime data ($2\nu2\beta$, EC, $\beta^-$) are then accurately 
reproduced by means of two free parameters $(g_{pp},\,g_A)$, resulting in an 
overconstrained parameter space.

\begin{figure}[t]
\begin{minipage}{17.5pc}
\includegraphics[height=3.1in,width=7cm]{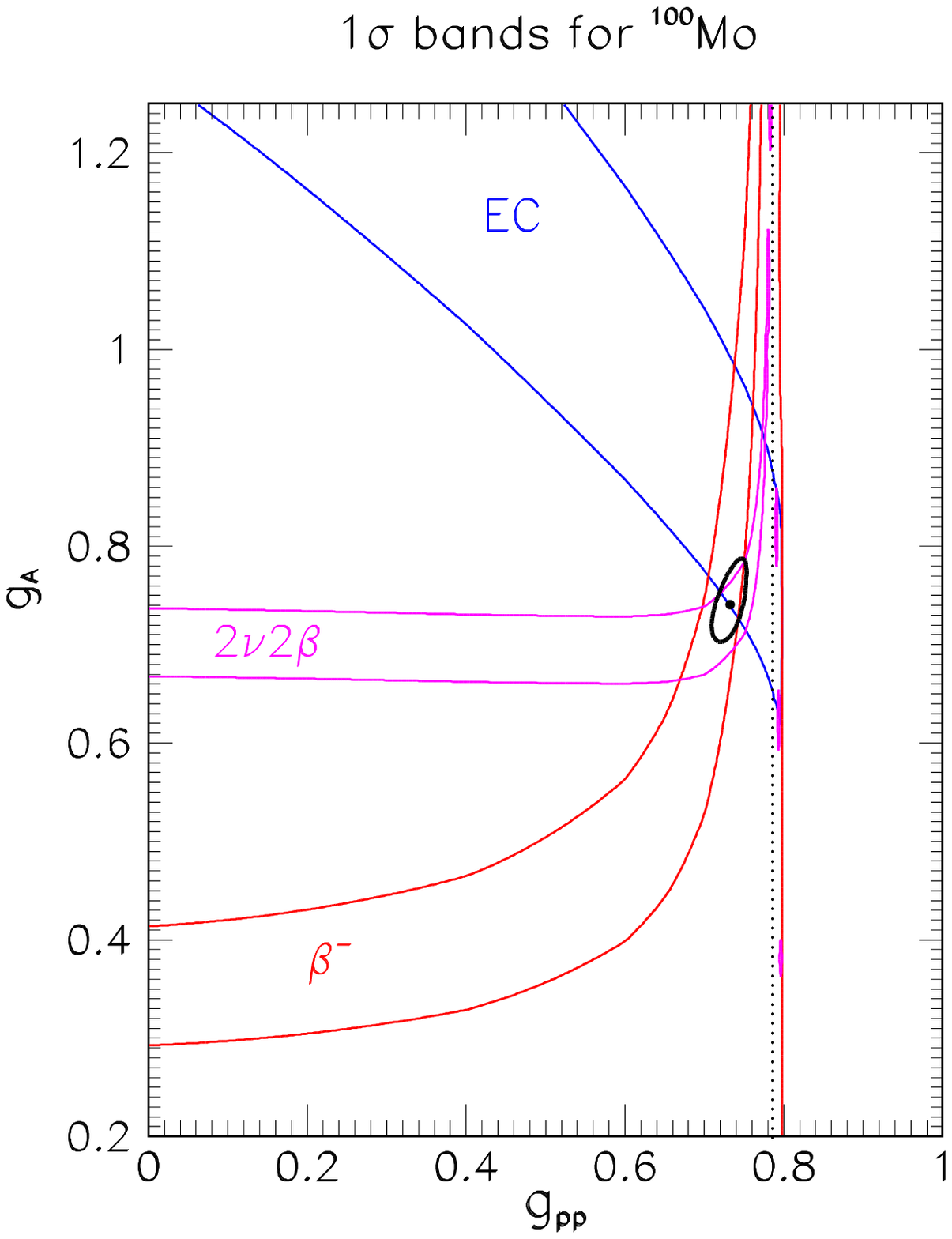}
\end{minipage}
\hspace{1pc}%
\begin{minipage}{17.5pc}
\includegraphics[height=3.1in,width=7cm]{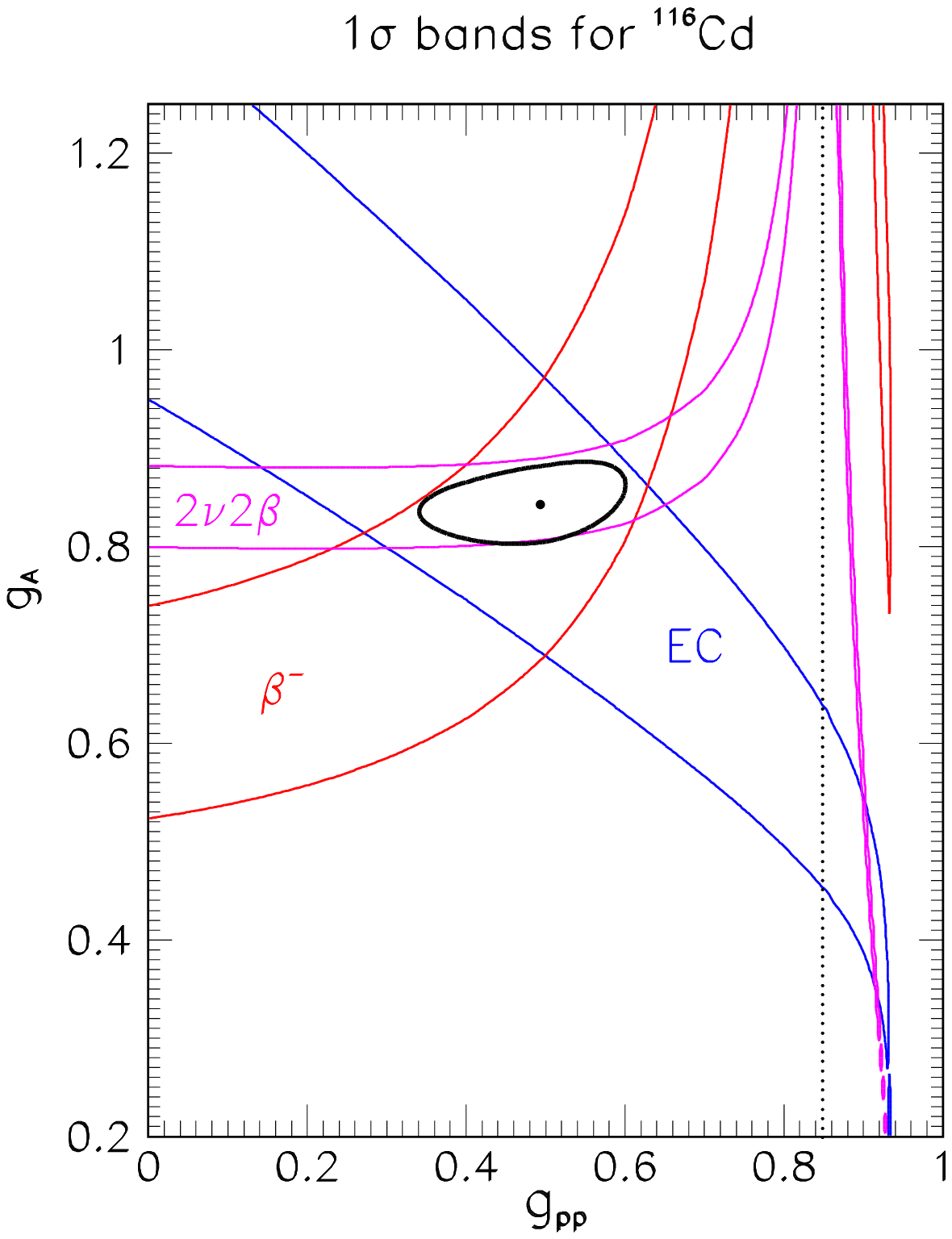}
\end{minipage} 
\caption{
Breakdown of individual constraints in the $(g_{pp},\,g_A)$ plane. The slanted bands corresponds to the regions allowed at $1\sigma$ level
(including experimental and theoretical errors) by $\beta^-$, EC, and $2\nu2\beta$ data. Their combination  is shown as a thick ellipse.
Left: $^{100}$Mo. Right: $^{116}$Cd. From Ref.~\protect\cite{nme}.}
\end{figure}

Overconstraining the $(g_{pp},~g_A)$ parameters is equivalent to
state that, in each of the $^{100}$Mo and $^{116}$Cd reference nuclei, our approach
provides one prediction which is experimentally verified. Figure~5 illustrates this statement
via the $1\sigma$ bands individually allowed by $\beta^-$, EC and $2\nu2\beta$ data for $^{100}$Mo and $^{116}$Cd. Any two
bands can be used to constrain $(g_{pp},~g_A)$ in a closed region (the ``prediction''), which is then 
crossed by the third independent band (the ``experimental verification''). 
As a consequence, the parametric
QRPA uncertainties induced by $(g_{pp},\,g_A)$ in the $0\nu2\beta$ decay
of these two nuclei are also significantly constrained~\cite{nme}.
We are planning a more systematic study, in order 
to extend this (or a similar) approach to other
nuclei of interest for $0\nu2\beta$ decay.

\section{Conclusions}

Since the atmospheric $\nu$ oscillation discovery 10 years ago,
important pieces of information are being slowly added to the puzzle of 
absolute $\nu$ masses. We have discussed the most recent 
oscillation and non-oscillation updates in the field, 
as presented at this NO-VE Workshop---and updated
after the recent  Neutrino~2008 Conference.
Oscillation parameters are robustly constrained,
and an intriguing indication for  $\theta_{13}>0$ appears to emerge.
Concerning non-oscillation observables, despite some recent
experimental and theoretical progress, a coherent picture remains 
elusive. In particular, the $0\nu2\beta$ claim is still under independent experimental scrutiny, and
it may be compatible 
or incompatible with the 
cosmological bounds, depending on data selection (especially Ly$\alpha$).
Reduction of nuclear matrix element uncertainties is also crucial
to improve the comparison of different $0\nu2\beta$ results. 
A confident assessment of the $\nu$ mass
scale will require converging evidence from at least two 
of the three observables $(m_\beta,\,m_{\beta\beta},\,\Sigma)$ within the
narrow limits allowed by oscillation data.

\section{Acknowledgments}
G.L.F., E.L., A.M., and A.M.R.\ acknowledge 
support by the Italian MIUR and INFN through the ``Astroparticle Physics'' 
research project, and by the EU ILIAS through the ENTApP project. 
A.P.\ is supported by MEC under the I3P program, by Spanish grants FPA2005-01269
and by European Commission network MRTN-CT-2004-503369 and
ILIAS/N6 RII3-CT-2004-506222. 
G.L.F. and E.L.\ thank Milla Baldo Ceolin for kind hospitality in Venice.

\medskip
\medskip\medskip\medskip
Note: For the sake of brevity, the following 
bibliography is essentially  limited to some of our recent papers. 
One can find therein relevant references, as well as credit to previous works,
about the vast phenomenology of neutrino masses and mixings.

\end{document}